\documentclass[aps,prl,twocolumn,floatfix,showpacs]{revtex4}
\usepackage{amsmath,amssymb,graphicx,bm}
\newdimen{\sGraphH}
\newsavebox{\GraphBox}
\savebox{\GraphBox}{\resizebox{!}{\sGraphH}{%
 \includegraphics
 {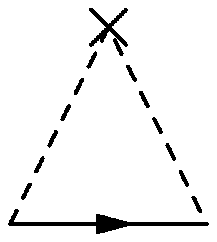}}}
\newdimen{\sGraphW}
\savebox{\GraphBox}{\resizebox{\sGraphW}{!}{%
 \includegraphics
 {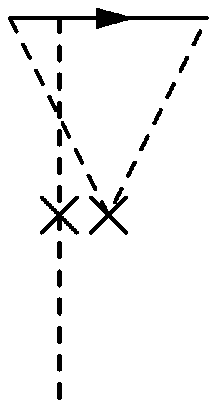}}}
\newdimen{\vGraphH}
\begin{document}
\title{Causality vs. Ward identity in disordered electron
systems}

\author{V.  Jani\v{s}} \author{ J. Koloren\v{c}}

\affiliation{Institute of Physics, Academy of Sciences of the Czech
  Republic, Na Slovance 2, CZ-18221 Praha 8, Czech Republic}
\email{janis@fzu.cz, kolorenc@fzu.cz}

\date{\today}


\begin{abstract}
We address the problem of fulfilling consistency conditions in solutions
for disordered noninteracting electrons. We prove that if we assume the
existence of the diffusion pole in an electron-hole symmetric theory we
cannot achieve a solution with a causal self-energy that would fully fit
the Ward identity.  Since  the self-energy must be causal, we conclude that
the Ward identity is partly violated in the diffusive transport
regime of disordered electrons. We explain this violation in physical
terms  and  discuss its consequences.
\end{abstract}

\maketitle 
Quantum charge transport in random media is generally a rather complex
problem. Unfortunately, even the simplest quantum tight-binding model of
disordered noninteracting electrons does not provide for an exact
solution. To find a complete solution of the Hamiltonian with random
energies amounts to the determination of the spectrum of the Hamiltonian
together with the spatial extension of the eigenfunctions. If the disorder
is sufficiently small the eigenfunctions are extended waves enabling
diffusion with a metallic conductivity. As shown by Anderson in his
seminal paper on quantum diffusion, Ref.~\onlinecite{Anderson58}, strong
randomness in the scattering potential can, however, destroy long-range
tails of the extended eigenfunctions and cause electron localization.
Localized eigenfunctions do not contribute to diffusion and drive the
system insulating. After years of intensive studies, full understanding of
Anderson localization has not yet been achieved and it still remains a big
challenge in condensed matter theory.

Although there are efficient mathematical tools to handle finite random
matrices,\cite{Mehta67} it is more practical to use configurationally
averaged Green functions to characterize solutions of random systems.
Green functions have proved useful in particular in the thermodynamic
limit where the volume and the number of particles are infinite. Exact
solutions are mostly elusive  and we have to resort to approximations.
A mean-field theory of the Anderson model of disordered electrons, the
coherent-potential approximation (CPA),\cite{Elliot74} contains only
extended states. Various extensions beyond the CPA are unable to include
localized eigenstates either.\cite{Gonis92} To take into account localized
states, one has to treat and approximate one- and two-electron averaged
Green functions simultaneously. The one-electron resolvent determines the
spectrum of the solution, while the two-electron resolvent determines the
spatial extension of the underlying eigenstates.

To arrive at a consistent solution with one- and two-electron averaged
Green functions we have to comply with a number of restrictive consistency
conditions reflecting analyticity and causality on the one side and
conservation laws on the other. First of all we have to warrant that the
eigenenergies of the random Hamiltonian remain real. This is the case if
the averaged resolvent $G(\mathbf{k},z)= \left\langle\langle\mathbf{k}|
\bigl[z\widehat{1} - \widehat{H}\bigr]^{-1}|\mathbf{k}
\rangle\right\rangle_{av}$ has no poles in the upper and lower half-planes
of complex energies~$z$. We denote here $|\mathbf{k}\rangle$ the
eigenfunctions of the nonrandom kinetic energy (Bloch waves), labelled by
momenta $\mathbf{k}$. The analytic behavior of the averaged resolvent is
expressed in the Kramers-Kronig relations between its real and imaginary
parts. If we use the Dyson equation to represent the averaged resolvent
via a self-energy $G(\mathbf{k},z) = \left[z -\epsilon(\mathbf{k}) -
\Sigma(\mathbf{k},z)\right]^{-1}$ we can write the Kramers-Kronig
relations for the self-energy
\begin{subequations}\label{eq:SE_KK}
 \begin{equation}
  \label{eq:SE_KK1}
  \Re \Sigma({\bf k},E+i\eta) = \Sigma_\infty + P\int_{-\infty}^{\infty}
   \frac{d E'}{\pi} \ \frac{\Im\Sigma({\bf
   k},E'+i\eta)}{E'-E}\ ,
\end{equation}
 \begin{equation}
\label{eq:SE_KK2} \Im \Sigma({\bf k},E+i\eta)= -P\int_{-\infty}^{\infty}
\frac{d E'}{\pi} \ \frac{\Re\Sigma({\bf k},E'+i\eta)}{E'-E}\ ,
\end{equation}
\end{subequations}
where $\Sigma_\infty$ is the constant part of the self-energy (its
high-energy limit).

When constructing two-particle functions we have to obey besides
analyticity also conservation laws expressed in Ward identities. Vollhardt
and W\"olfle proved the following integral Ward identity in the
Anderson model of disordered electrons\cite{Vollhardt80b}
\begin{multline}
  \label{eq:VWW_identity}
  \Sigma({\bf k}_+,z_+) - \Sigma({\bf k}_-, z_-) = \frac 1N \sum_{{\bf
      k}'}\Lambda^{eh}_{{\bf k}{\bf k}'}(z_+,z_-;{\bf q})\\ \times
\left[G({\bf       k}'_+,z_+) - G({\bf k}'_-, z_-) \right] \ ,
\end{multline}
where $\Lambda^{eh}$ stands for the irreducible vertex in the electron-hole
scattering channel, ${\bf k}_\pm={\bf k}\pm{\bf q}/2$, and $z_\pm=E\pm
\omega \pm i\eta$, $\eta>0$. The original proof of identity
\eqref{eq:VWW_identity} did not reveal the conservation law reflected by
this identity, but it was shown later that Eq.~\eqref{eq:VWW_identity} is
a consequence of probability conservation in the Hilbert space of Bloch
waves.\cite{Janis01b}

Identity \eqref{eq:VWW_identity} plays an important
role in physics of disordered electrons, since it sets up a
condition for the existence of the diffusion pole in the density-density
correlation function.\cite{Vollhardt92,Janis02a} The diffusion pole in
the electron-hole correlation function is a hallmark for the diffusive
character of charge transport in disordered systems.

Neither Eq.~\eqref{eq:VWW_identity} nor Eq.~\eqref{eq:SE_KK} are
automatically guaranteed in quantitative treatments of disordered
electrons. In fact, different conditions must be fulfilled to comply with
either Eq.~\eqref{eq:SE_KK} or Eq.~\eqref{eq:VWW_identity}. These
conditions can be contradictory in specific situations and it need not be
always possible to comply with both equations simultaneously. It is the
aim of this Letter to materialize such a situation in models of disordered
noninteracting  electrons. The principal result
of this Letter is the following\\
\textbf{Assertion}: \emph{If we assume electron-hole symmetry in the vertex
functions and the existence of a diffusion pole in the low-energy limit of
the electron-hole correlation function having the explicit form $2\pi n_F/
[-i\omega + Dq^2]$, it is impossible to accommodate
Eq.~\eqref{eq:VWW_identity} for $z_+- z_- \gtrless 0$ with causality of
the self-energy, i.~e., Eqs.~\eqref{eq:SE_KK}.}

Before we proceed with a  proof we first briefly discuss the assumptions
in this assertion  and illustrate in simple terms how a conflict between
causality and the Ward identity \eqref{eq:VWW_identity} emerges.

To see how causality of the self-energy can be traced in the
electron-hole irreducible vertex we reduce the complex frequency
difference in  Eq.~\eqref{eq:VWW_identity} to a purely imaginary value,
i.~e. $z_+-z_-=2i\eta$ and set $q=0$ to obtain  a special case of the Ward
identity
\begin{multline}
\label{eq:SE_vertex_im}
\Im\Sigma({\bf k},E+i\eta)\\ =\frac 1N\sum_{{\bf
k}'}\Lambda^{eh}_{{\bf k}{\bf k}'}(E+i\eta,E-i\eta)\Im G({\bf k}',E+i\eta).
\end{multline}
We have denoted $\Lambda^{eh}_{{\bf k}{\bf k}'}(z_+,z_-) \equiv
\Lambda^{eh}_{{\bf k}{\bf k}'}(z_+,z_-;\mathbf{0})$. Causality of the
self-energy (negative semi-definiteness of its imaginary part) can then be
controlled by positive semi-definiteness of the self-adjoint part of the
electron-hole vertex appearing on the r.h.s. of
Eq.~\eqref{eq:SE_vertex_im}. A causal self-energy can be determined from
a given positive semi-definite electron-hole irreducible vertex. One uses
the Ward identity to resolve the imaginary part of the self-energy from
the electron-hole vertex and completes it with Eq.~\eqref{eq:SE_KK1} to
determine the real part of the self-energy.

The self-energy calculated from the electron-hole irreducible vertex via
Eqs.~\eqref{eq:SE_vertex_im} and \eqref{eq:SE_KK1} is causal and uniquely
determined. We can warrant fulfillment of the Ward
identity for imaginary energy differences, i.~e., $\omega=0$,  in this
construction. Behavior of the electron-hole vertex for $\omega\neq0$ is
irrelevant for the self-energy. Whether the self-energy from
Eq.~\eqref{eq:SE_vertex_im} satisfies the Ward identity also for
real-energy differences remains to be checked.
 
We know that the mean-field solution (CPA) obeys the Ward identity for
arbitrary difference of complex energies. It is due to locality of the CPA
electron-hole irreducible vertex.\cite{Note0} To find out whether
Eq.~\eqref{eq:VWW_identity} can generally be fulfilled we hence have to
analyze nonlocal electron-hole irreducible vertices.

The simplest nonlocal contributions to the electron-hole vertex are crossed
diagrams. The self-energy is causal when the electron-hole irreducible
vertex is a (positive) sum of products $GG^*$. Perturbation
theory  up to second order for the electron-hole vertex with this property
reads
\begin{equation}\label{eq:vertex-causal}
\begin{split}
\hbox{
  \vbox to \vGraphH{\vfill\hbox{$\Lambda_{C}^{eh}=$}\vfill}
  \vbox{\hbox{\resizebox{!}{\vGraphH}{%
  \includegraphics{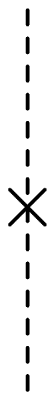}}}}
  \vbox to \vGraphH{\vfill\hbox{+}\vfill}
  \vbox{\hbox{\resizebox{!}{\vGraphH}{%
 \includegraphics
 {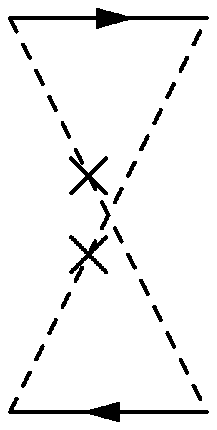}}}}
}
\end{split}
\end{equation}
where the dashed line stands for the local impurity potential
$\lambda=c\left(\langle V^2\rangle_{av} - \langle V\rangle^2_{av}\right)$
and the oriented solid lines are renormalized electron (retarded, upper
line) and hole (advanced, lower line) averaged propagators. The
approximate self-energy constructed from the vertex $\Lambda^{eh}_C$ via
Eq.~\eqref{eq:SE_vertex_im} and Eq.~\eqref{eq:SE_KK1} is causal. It does
not, however, obey the Ward identity \eqref{eq:VWW_identity} as can easily
be checked by means of its proof.\cite{Janis02a}

To comply with the Vollhardt-W\"olfle identity
\eqref{eq:VWW_identity} up to  second order we have to supplement
the simplest (maximally) crossed diagram with two further diagrams of the
same order. That is, instead of Eq.~\eqref{eq:vertex-causal} we have to
consider the following electron-hole vertex
\begin{subequations}
\begin{equation}\label{eq:Lambda_WI}
\begin{split}
\hbox{
  \vbox to \vGraphH{\vfill\hbox{$\Lambda_{WI}^{eh}=$}\vfill}
  \vbox{\hbox{\resizebox{!}{\vGraphH}{%
   \includegraphics
   {vertexborngraph.eps}}}}
  \vbox to \vGraphH{\vfill\hbox{+}\vfill}
  \vbox{\hbox{\resizebox{!}{\vGraphH}{%
   \includegraphics
   {vertex1crossgraph.eps}}}}
  \vbox to \vGraphH{\vfill\hbox{+}\vfill}
  \vbox{\hbox{\resizebox{!}{\vGraphH}{%
   \includegraphics
{vertex1crossirrelagraph.eps}}}}
  \vbox to \vGraphH{\vfill\hbox{+}\vfill}
  \vbox{\hbox{\resizebox{!}{\vGraphH}{%
   \includegraphics
{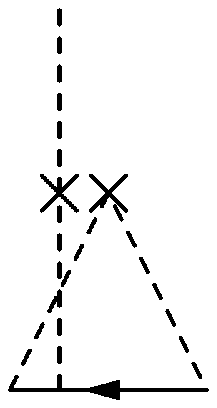}}}}
  \vbox to \vGraphH{\vfill\hbox{.}\vfill}
}
\end{split}
\end{equation}
We can find a formal diagrammatic solution for the
self-energy corresponding to the vertex $\Lambda^{eh}_{WI}$:
\begin{equation}\label{eq:Sigma_WI}
\begin{split} 
\hbox{
  \vbox to \sGraphH{\vfill\hbox{$\Sigma_{WI}=$}\vfill}
  \vbox{\hbox{\resizebox{!}{\sGraphH}{%
    \includegraphics
     {selfenborngraph.eps}}}}
  \vbox to \sGraphH{\vfill\hbox{+}\vfill}
  \vbox{\hbox{\resizebox{!}{\sGraphH}{%
  \includegraphics
  {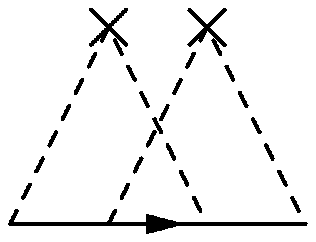}}}}
  \vbox to \sGraphH{\vfill\hbox{.}\vfill}
}
\end{split}
\end{equation}
\end{subequations}
The Ward identity holds for $\Sigma_{WI}$ and $\Lambda^{eh}_{WI}$, but the
self-energy is not causal. The imaginary part $\Im\Sigma^R_{WI}$ becomes
positive for specific energies and disorder strengths. It hence does not
fulfil the Kramers-Kronig relation \eqref{eq:SE_KK1} and consequently it
cannot be obtained from Eq.~\eqref{eq:Lambda_WI} via
Eq.~\eqref{eq:SE_vertex_im}. To amend this we have to add an infinite
series of compensating diagrams (from the so called traveling clusters) so
that the resulting self-energy representation generates a Herglotz
function.\cite{Mills78}

The situation cannot be remedied by adding higher-order maximally crossed
diagrams, i.~e., diagrams with scatterings only between the
electron and the hole where each impurity line crosses all other impurity
potentials. With each maximally crossed diagram of order $n$,
generating a causal contribution to the self-energy, we have to add $n$
additional compensating non-causal diagrams.\cite{Janis02a} The corrected
irreducible vertex, compatible with the Ward identity, generates a
non-Herglotz self-energy and the sum of diagrams is not causal at any order
of the expansion.

It is impossible to find a \emph{diagrammatic} representation of a causal
self-energy obeying Eq.~\eqref{eq:VWW_identity} for real frequency
differences and containing a sum of maximally crossed diagrams.
Inconsistency between causality and the Ward identity in perturbation
theory originates in these nonlocal diagrams. To prove incompatibility of
causality with the Ward identity nonperturbatively we need an infinite sum
of maximally crossed diagrams producing a singularity in the electron-hole
irreducible vertex. This singularity, the Cooper pole, is generally an
image of the diffusion pole in the full electron-hole vertex. The
diffusion pole in the low-energy limit of the electron-hole correlation
function was proved to have exactly the form assumed in the
Assertion.\cite{Janis02a}

\textbf{Proof of the Assertion}. The diffusion pole in the electron-hole
correlation function is a residue of a pole in a more fundamental
full two-particle vertex (Green) function $\Gamma_{\mathbf{k}\mathbf{k}'}
(z_+,z_-;\mathbf{q})$. We denote $\Lambda^{eh}$ and $\Lambda^{ee}$
the irreducible vertices in the electron-hole and the electron-electron
scattering channels characterized by antiparallel and parallel propagator
lines, respectively. Let further $I$ be a vertex irreducible in both
channels. The following identity $\Lambda^{eh} + \Lambda^{ee} = I +
\Gamma$ then holds. Since the vertex $I$ is regular, the pole from the
full vertex $\Gamma$ must appear at least in one of the irreducible
vertices. Due to the electron-hole symmetry,
$\Lambda^{eh}_{\mathbf{k}\mathbf{k}'}(z_+,z_-;\mathbf{q}) =
\Lambda^{ee}_{\mathbf{k}\mathbf{k}'}(z_+,z_-;-\mathbf{q} - \mathbf{k}
- \mathbf{k}')$ both irreducible vertices share the same
singularity.

The explicit singular low-energy asymptotics of the electron-hole
irreducible vertex $\Lambda^{eh}$ can be obtained from the sum of all
maximally crossed diagrams. At zero temperature for $q=0$ and $|\mathbf{k}
+ \mathbf{k}'|\to0$ this sum yields
\begin{equation}\label{eq:Cooper-pole}
\Lambda^{sing}_{\mathbf{k}\mathbf{k}'}(z_+,z_-) \doteq \frac{2\pi
n_F\lambda} {-i\Delta z\ \text{sign}\left(\Im\Delta z\right) + D(\mathbf{k}
+ \mathbf{k}')^2}\ .
\end{equation}
We have denoted $\Delta z = z_+ - z_-$, $n_F$ the density of states at the
Fermi energy, and $D$ the diffusion constant. This form of the low-energy
behavior holds also for stronger scatterings on impurities. Multiple
scatterings only renormalize characteristic parameters $\lambda$ and $D$
of the singular part of the electron-hole vertex for $\Delta
z\to0$.\cite{Vollhardt92} Equation~\eqref{eq:Cooper-pole} is a
generalization of the diffusion pole from real to complex frequency
differences.
 
To demonstrate incompatibility of the Ward identity with causality of the
self-energy we define a real-frequency function
\begin{subequations}
\begin{equation} \label{eq:SE_difference}
\Delta W(\omega) = \frac{1}N \sum_{\mathbf{k}}
[\Sigma^R(\mathbf{k},E-\omega) - \Sigma^R(\mathbf{k},E +\omega)]\ .
\end{equation}
From the Ward identity we obtain its representation via the irreducible
electron-hole vertex
\begin{multline}\label{eq:vertex-difference}
\Delta W(\omega) = \frac {-1}{N^2}\ \sum_{\mathbf{k}\mathbf{k}'}\left\{2i
\left[\Lambda^{RA}_{\mathbf{k} \mathbf{k}'}(E+\omega,E) \right.\right. \\
\left.\left. - \Lambda^{RA}_{\mathbf{k} \mathbf{k}'}(E-\omega,E)\right]
\Im G^R(\mathbf{k}',E) + \Lambda^{RA}_{\mathbf{k}
\mathbf{k}'}(E+\omega,E)\right. \\ \left. \times
\left[G^R_{\mathbf{k}'}(E+\omega) - G^R_{\mathbf{k}'}(E)\right] -
\Lambda^{RA}_{\mathbf{k} \mathbf{k}'}(E-\omega,E) \right. \\
\left.\times\left[G^R_{\mathbf{k}'}(E-\omega) - G^R_{\mathbf{k}'}(E)\right]
\right\} \ .
\end{multline}
\end{subequations}
We have used superscripts $R,A$ to denote the complex half-planes of the
energy arguments (upper, lower) from which the real value is approached.
Analyticity of the self-energy as a function of energy $E$ inside the
energy bands induces analyticity of the function $\Delta W(\omega)$,
i.~e., all its derivatives are continuous functions and bounded at
$\omega=0$.

The vertex function $\Lambda^{RA}$ is not, however, analytic and its
nonanalyticity is transferred to $\Delta W(\omega)$. It is straightforward
to evaluate $\Delta W^{sing}(\omega)$, the leading singular part of $\Delta
W(\omega)$ in the limit $\omega\to0$. We rescale the momenta $q\to x=
q\sqrt{D/|\omega|}$ in $\Lambda^{RA}$ and perform momentum integration in
a $d$-dimensional space. The leading singular behavior is
determined by the small-momentum asymptotics of the first term on the
r.h.s. of Eq.~\eqref{eq:vertex-difference} with the singular  electron-hole
vertex from Eq.~\eqref{eq:Cooper-pole} for $\eta=0^+$. We explicitly obtain
\begin{multline}\label{eq:nonanalyticity}
\Delta W^{sing}_d(\omega)  \thickapprox K \lambda n_F^2 \\ \times
\begin{cases}
 \frac 1{\omega} \left|\frac{\omega}{Dk_F^2}\right|^{d/2}\
 & \text{for $d\neq4l$},\\
\frac 1{\omega} \left|\frac{\omega}{Dk_F^2}\right|^{d/2}\  \ln\left|\frac
{Dk_F^2}\omega\right| & \text{for $d=4l$},
\end{cases}
\end{multline}
where $K$ and $k_F$ are a dimensionless constant of order one and the Fermi
momentum, respectively.\cite{Note1} Equation~\eqref{eq:SE_difference},
however, dictates  that $\Delta W^{sing}_d(\omega)=0$ almost
everywhere inside the energy bands. We hence have a contradiction to the
premise of validity of the Ward identity Eq.~\eqref{eq:VWW_identity} for
real frequency differences. \textbf{Q.~E.~D.}

We can see from Eq.~\eqref{eq:nonanalyticity} that the nonanalyticity in
$\Delta W(\omega)$ strongly depends on the spatial dimension $d$. It has a
square root character in odd dimensions, contains a sign function in
dimensions $d=2(2l+1)$ and a logarithm in $d=4l$. The nonanalytic behavior
is smoothened with increasing dimensionality of the system, i.~e., the
higher the dimension the higher is the derivative of $\Delta W(\omega)$
displaying an (infinite) jump at $\omega=0$. It is the $[(d-2)/2]$th
derivative, where $[x]$ is the smallest integer greater than or equal to
$x$. As expected, there is no nonanalyticity in the mean-field,
$d=\infty$, limit of the electron-hole vertex and one cannot use a $1/d$
expansion around the CPA  in two-particle functions.

The nonanalyticity in $\Delta W(\omega)$ demonstrates that we cannot
simultaneously keep causality and the Ward identity.
Causality of the self-energy (reality of eigenenergies of the random
Hamiltonian) cannot be broken. We hence are forced to  admit a violation of
the Ward identity for real frequency differences $z_+-z_-=2\omega\neq0$ in
theories with a causal self-energy and the Cooper pole in the electron-hole
irreducible vertex.

Giving up the Ward identity has quantitative consequences for the diffusive
behavior of disordered electrons. Most importantly, the asymptotic form of
the diffusion/Cooper pole deviates from Eq.~\eqref{eq:Cooper-pole}. This
deviation in the low-energy asymptotics of the electron-hole correlation
function can be explicitly manifested in high spatial dimensions. There we
expect only a weak, negligible momentum dependence of the self-energy as
well as of integrals of the vertex function. The Cooper singularity in the
vertex function is sufficiently smoothened by momentum integration. If we
introduce a parameter
\begin{multline}\label{eq:lambda_diff}
A = -\frac{\partial}{\partial\omega}\bigg[\Sigma^R(\mathbf{k},E+\omega)
 - \frac 1N \sum_{{\bf k}'} \Lambda^{RA}_{\mathbf{k}
\mathbf{k}'} (E+\omega, E ) \\  \times
\left(G^R(\mathbf{k}',E+\omega) -
G^A(\mathbf{k}',E)\right)\bigg]_{\omega=0}
\end{multline}
we can represent the leading low-energy asymptotics of the
zero-temperature electron-hole correlation function in high spatial
dimensions as
\begin{multline}
\label{eq:Phi_diffusion} \Phi^{RA}_E({\bf q},\omega)= \frac
1{N^2}\sum_{\mathbf{k}\mathbf{k}'} G^{RA}_{\mathbf{k}\mathbf{k}'}
(E+\omega,E;\mathbf{q})\\  \approx \frac {2\pi n_F}{-i\omega(1+A) + Dq^2} .
\end{multline}

The parameter  $A$ increases with the disorder strength. In
particular, it is the imaginary part of the frequency derivative of the
electron-hole vertex, $\Im\partial \Lambda^{eh}/\partial\omega$, that drives
the system toward the Anderson localization transition, where it eventually
diverges. The weight of the diffusion pole $n_F/(1+A)$, expressing the
number of electron states available for diffusion, decreases when
approaching the localization transition as well as an effective diffusion
constant $D/(1+A)$. They both vanish at the Anderson localization
transition. There is hence no diffusion pole in the localized phase.

Violation of the Ward identity can be consistently explained and does not
actually contradict any fundamental physical law.\cite{Note2} The Ward
identity holds only within the Hilbert space spanned by extended Bloch
waves, the eigenstates of the nonrandom kinetic-energy
operator.\cite{Janis02a} Violation of the Ward identity then expresses the
fact that we cannot construct a single Hilbert space for all
configurations of the random potential. There are new, disorder-induced
and configuration-dependent localized states \emph{orthogonal} in the
thermodynamic limit to Bloch (plane) waves. The statistical weight of
these states is not negligible and hence the state vector of the electron
in random media has a non-zero projection onto a space of localized states
at any disorder strength. That is why the number of electrons in the space
of extended states is not conserved when the disorder strength increases.
The weight of the localized states in the decomposition of the electron
state vector increases and reaches its saturation at the Anderson
localization transition, where the state vector becomes orthogonal to
Bloch waves.

To conclude, we proved in this Letter that the Cooper pole,
Eq.~\eqref{eq:Cooper-pole}, in the electron-hole vertex leads in systems
with disordered electrons to deviations from the Ward identity. The
nonanalytic behavior due to the Cooper pole in the electron-hole vertex is
not sufficiently smeared out by momentum integration in the Ward identity
to produce a self-energy analytic in its energy argument. The observed
deviation from the Ward identity in causal theories is explained by
nonconservation of electron states available for diffusion, i.e.,
plane waves. Due to the existence of disorder-induced localized states,
the electrons may escape from the Hilbert space of extended states into the
space of localized ones, orthogonal to Bloch waves.

The research on this problem was carried out within a project AVOZ1-010-914
of the Academy of Sciences of the Czech Republic and supported in part by
Grant No. 202/01/0764 of the Grant Agency of the Czech Republic. The
authors thank V. \v{S}pi\v{c}ka for discussions on quantum coherence
and diffusion and VJ thanks D. Vollhardt for hospitality at the University
of Augsburg and fruitful discussions on Ward identities and
self-consistent theories of Anderson localization.

\end{document}